# Directions In Optical Angular Momentum

Matt M. Coles and David L. Andrews*

*School of Chemistry, University of East Anglia, Norwich NR4 7TJ, United Kingdom*
*\* To whom correspondence should be addressed. E-mail: d.l.andrews@uea.ac.uk*

**Abstract.** For paraxial and non-paraxial light, numerous measures of electromagnetic attribute are expressible in terms of photon annihilation and creation. Accordingly, energy, angular momentum and chirality measures acquire a consistent interpretation. From the photonic nature of light, it emerges that an infinite hierarchy of spin-type measures depend on a difference in number operators for modes of opposing helicity, pure circular polarization giving maximal values. Measures of orbital angular momentum are determined by a sum of corresponding number operators. By analysis of electric and magnetic features in the reflection of circular polarizations, regions of prominent chiral interaction in the interference are identified.

It has recently been pointed out [1] that a classical text-book treatment of plane electromagnetic waves, with transverse electric and magnetic field vectors, delivers vanishing angular momentum density along the axis of a beam in the direction of propagation. Any quantum treatment of circularly polarized light, however, results in an intrinsic angular momentum of $\hbar$ per photon, independent of photon energy [2]. The torque exerted about the beam axis in Beth's historic study provided the first experimental indication of a longitudinal angular momentum component [3].

The relationships between the various measures of optical angular momentum can first be addressed by noting the well-known decomposition [4] of total electromagnetic angular momentum into spin and orbital components. In general, the angular momentum **J** of the electromagnetic field can be defined as:

$$\mathbf{J} = \varepsilon_0 \int d^3\mathbf{r}\, \mathbf{r} \times (\mathbf{E} \times \mathbf{B}), \quad (1)$$

where $\varepsilon_0$ is the vacuum permittivity and **E**, **B**, respectively, are the electric and magnetic fields implicitly evaluated at position **r**. This *total* angular momentum can be recast as the sum of the following terms:

$$\mathbf{L} = \varepsilon_0 \int d^3\mathbf{r}\, E_i (\mathbf{r} \times \nabla) \cdot A_i, \quad (2)$$

$$\mathbf{S} = \varepsilon_0 \int d^3\mathbf{r}\, (\mathbf{E} \times \mathbf{A}), \quad (3)$$

**L** signifying the *orbital* angular momentum for the field, and **S** the *spin* angular momentum, where the Einstein summation convention is used and **A** is the electromagnetic vector potential. It has been recognized that this separation, through the explicit involvement of the vector potential, is gauge-dependent. However, using the paraxial approximation, or requiring that the **A** field is evaluated in the Coulomb gauge, allows justification of the split into Eqs (2), (3) [5].

In a quantum field representation the classical **A**, **E** and **B** fields are promoted to Hilbert space operators; the electromagnetic vector potential is given by:

$$\mathbf{A} = \sum_{\mathbf{k},\eta} \left( \frac{\hbar}{2\varepsilon_0 c k V} \right)^{\frac{1}{2}} \left\{ \mathbf{e}^{(\eta)}(\mathbf{k}) a^{(\eta)}(\mathbf{k}) e^{i(\mathbf{k}\cdot\mathbf{r})} + h.c. \right\}, \quad (4)$$

where $\hbar$ is the reduced Planck constant, $V$ is the quantization volume, while $\mathbf{e}^{(\eta)}(\mathbf{k})$ and $a^{(\eta)}(\mathbf{k})$ are the polarization vector and photon annihilation operator respectively, corresponding to a mode with polarization $\eta$ and wave-vector **k**; *h.c.* represents Hermitian conjugate [6]. Here, and in the following, we only explicitly exhibit operator form (by a carat) where necessary to preclude ambiguity. Using $\mathbf{B} = \nabla \times \mathbf{A}$ and $\mathbf{E} = -\partial \mathbf{A}/\partial t$, the **E** and **B** fields can be derived from Eq. (4). Electromagnetic measures that depend on these fields and potentials also become operators through second quantization.

All of the sought measures of freely propagating optical radiation have to be delivered as expectation values, corresponding to observations that leave the system unchanged. It can be asserted that the associated Hermitian operators emerge in terms comprising equal numbers of annihilation and creation operators. For example, in the paraxial approximation the components of the electromagnetic stress-energy 4-tensor – the energy density, the Poynting vector and the Maxwell stress tensor – are all bilinear in the **E** and **B** fields. For paraxial radiation it is evident that the evaluation of each of these quantities yields a result with only terms containing precisely one annihilation and creation operator. To address the case of non-paraxial beams we use instead an exact classical solution [7] for the electric field vector

$$\mathbf{E}(x) = e^{il\phi} \int_0^k d\kappa E(\kappa) e^{ik_z z} \times \left\{ \left( \alpha_u e_x + \alpha_v e_y \right) J_l(\kappa\rho) \right.$$
$$\left. + e_z \frac{\kappa}{2k_z} \left[ (i\alpha_u - \alpha_v) e^{-i\phi} J_{l-1}(\kappa\rho) - (i\alpha_u + \alpha_v) e^{i\phi} J_{l+1}(\kappa\rho) \right] \right\},$$
(5)

expanded in terms of Bessel functions $J_l(\kappa\rho)$, where $l$ is the topological charge, $k_z = (k^2 - \kappa^2)^{1/2}$ and the complex constants $\alpha_u, \alpha_v$ satisfy $|\alpha_u|^2 + |\alpha_v|^2 = 1$. Recognizing that $\alpha_u$ and $\alpha_v$ relate to orthogonal plane polarizations, we can promote them to the operator status through:

$$\alpha_u \to \frac{1}{\sqrt{2}} \left[ a^{(L)}(\mathbf{k}) + a^{(R)}(\mathbf{k}) \right],$$

$$\alpha_v \to \frac{i}{\sqrt{2}} \left[ a^{(L)}(\mathbf{k}) - a^{(R)}(\mathbf{k}) \right],$$

where L/R denote left and right circular polarization, allowing Eq. (5) to be expressed in the required quantum formalism. Inspection then reveals that all terms are linear in $a^{(\eta)}(\mathbf{k})$ and $a^{\dagger(\eta)}(\mathbf{k})$, verifying that even for non-paraxial light, the observables – being bilinear in the electric and magnetic fields – invoke equal numbers of annihilation and creation operators. A legitimate question is whether a free-space electromagnetic measure based on unequal numbers of annihilation and creation operators can represent any physically meaningful quantity.

In quantum field theory, a product of field components (equally, in consequence, a corresponding product of field annihilation and creation operators), is in normal order when all creation operators are to the left of each annihilation operator [8]. Wick's Theorem states that any string of annihilation and creation operators can be decomposed into terms that are all in normal order. Significantly, the difference between the number of annihilation and creation operators is a constant for all terms in the decomposition. Such definitions, and the following arguments, apply equally well to analytical functions with arguments containing strings of $a^{(\eta)}, a^{\dagger(\eta)}$, since such functions admit power series expansions [9]. An operator $\hat{Q}$, corresponding to a conjectured electro-magnetic observable with terms containing potentially unequal numbers of annihilation and creation operators, transforms into a series of terms each with the form

$$\underbrace{a^{\dagger(\eta)} \ldots a^{\dagger(\eta)}}_{r} \underbrace{a^{(\eta)} \ldots a^{(\eta)}}_{s},$$

where $r$ and $s$ are the number of times each operator appears and $r$–$s$ is a constant dictated by $\hat{Q}$. Taking the expectation value of the terms of such an operator over a number (Fock) state delivers

$$\langle n | \underbrace{a^\dagger \ldots a^\dagger}_{r} \underbrace{a \ldots a}_{s} | n \rangle = \frac{n!^2}{(n-r)!(n-s)!} \langle n-r | n-s \rangle,$$

which gives a non-vanishing result only when $r = s$. Thus, in evaluating the observables of number states, the Hermitian operators acting on them must embody an equal number of lowering and raising operators.

To extend the argument, we turn attention to over-complete sets whose elements are not orthogonal, taking the familiar example of optical coherent states. Taking the expectation of the terms in $\hat{Q}$ gives

$$\langle \alpha | \underbrace{a^\dagger \ldots a^\dagger}_{r} \underbrace{a \ldots a}_{s} | \alpha \rangle = \bar{\alpha}^r \alpha^s,$$

where the eigenvalue $\alpha$ is a complex number; the result no longer vanishes for $r \neq s$. However, making explicit the time dependence in the raising and lowering operators, as in the interaction representation,

$$a^{(\eta)}(\mathbf{k}, t) = a^{(\eta)}(\mathbf{k}, 0) e^{-i\omega t}, \quad a^{\dagger(\eta)}(\mathbf{k}, t) = a^{\dagger(\eta)}(\mathbf{k}, 0) e^{i\omega t},$$

shows that a normally ordered string of such operators will contain an oscillating phase factor if the numbers of each operator are unbalanced – leading to a zero expectation value for $\hat{Q}$.

The issue discussed above informs recent deliberations concerning rediscovered measures of electromagnetic helicity, their physical meaning and relation to optical angular momentum [1, 10-14]. In previous work [12] we have shown that all of these quantities, given quantum operator status, display a dependence on the difference between number operators for optical modes of opposing helicity – a special case being when left and right handed modes are used as the basis, $\hat{N}^{(L)}(\mathbf{k}) - \hat{N}^{(R)}(\mathbf{k})$. It has also been shown by Cameron, Barnett and Yao [15] that an *optical chirality density* and its associated flux can be obtained from the electromagnetic helicity and spin angular momentum operators, by replacing every appearance of $\mathbf{A}$ and $\mathbf{C}$, the magnetic and electric vector potentials, with their curls, $\nabla \times \mathbf{A} = \mathbf{B}$ and $\nabla \times \mathbf{C} = -\mathbf{E}$. Moreover, the analysis showed that for any conserved electromagnetic quantity one can replace each appearance of fields with their curls, or curls of curls, and so on, generating an infinite list of related conserved quantities. It is now shown that starting with the electromagnetic helicity and spin operators,

$$H = \frac{1}{2} \int d^3 \mathbf{r} \left( \mathbf{A} \cdot \mathbf{B} - \mathbf{C} \cdot \mathbf{E} \right), \quad (6)$$

$$\mathbf{S} = \varepsilon_0 \frac{1}{2} \int d^3\mathbf{r} \left( \mathbf{E} \times \mathbf{A} + \mathbf{B} \times \mathbf{C} \right), \quad (7)$$

and repeatedly taking curls, creates families of helicity-type and spin-type measures, all proportional to differences between the populations of optical modes with opposing helicity. To complete the picture, it is observed that helicity and spin form a continuity equation,

$$\frac{\partial}{\partial t} H + \nabla \cdot \mathbf{S} = 0,$$

which is mirrored by a continuity equation for the 'higher order' optical chirality density and associated flux. Moreover, all helicity-type and spin-type measures have a conservation equation in their respective generations. Taking each successive curl of $\mathbf{A}$, Eq. (4), has the effect of multiplying the result by $ik$, changing the respective signs of the positive and negative frequency terms, and exchanging the polarization vectors in a 2-cycle,

$$\mathbf{e}^{(1/2)} \to \mathbf{k} \times \mathbf{e}^{(1/2)} \to \mathbf{e}^{(1/2)} \to \ldots$$

For greater generality, it is expedient to proceed with an analysis based on an arbitrary, appropriately chosen basis pair of polarization states. Any acceptable pair corresponds to diametrically opposing points on the traditional Poincaré sphere [16]. With a general polarization vector $\mathbf{e}_1$, characterized by angular coordinates $\theta$ and $\phi$, the counterpart basis vector $\mathbf{e}_2$ is generated according to the following prescription:

$$\left. \begin{array}{l} \mathbf{e}^{(1)} = \sin\theta \hat{\mathbf{i}} + e^{i\phi} \cos\theta \hat{\mathbf{j}} \\ \mathbf{e}^{(2)} = \cos\theta \hat{\mathbf{i}} - e^{i\phi} \sin\theta \hat{\mathbf{j}} \end{array} \right\}, \quad (8)$$

signifying geometrically opposite points on the sphere [17]. Any basis of this form satisfies the orthogonality condition, $\mathbf{e}^{(n)} \cdot \overline{\mathbf{e}}^{(m)} = \delta_{nm}$. Thus each of the helicity-type measures will always involve the scalar product of the polarization vector with the complex conjugate of the corresponding $\hat{\mathbf{k}} \times \mathbf{e}^{(1/2)}(\mathbf{k})$. Analysis of Eq. (8) shows that the polarization vectors, with the addition operation, form a mathematical group, resulting in the well-known observation that different proportions of left- and right- handed light produce an elliptical polarization state. Enacting the products described in Eqs (6), (7), and the sum over the polarization basis prescribed in the mode expansions, it is readily verified that terms involving

$$\mathbf{e}^{(1/2)}(\mathbf{k}) \cdot \overline{\left[ \hat{\mathbf{k}} \times \mathbf{e}^{(2/1)}(\mathbf{k}) \right]},$$

vanish. Thus, the remaining terms are those in which the $\mathbf{e}$-vectors in the product correspond to identical polarization states. As a consequence, the non-zero terms contain $a^{\dagger(1/2)}(\mathbf{k}) a^{(1/2)}(\mathbf{k}) = \hat{N}^{(1/2)}(\mathbf{k})$, with the polarization state matching that of the polarization vector. Finally, using the generalized Poincaré sphere basis exhibited in Eq. (8), the form of the polarization product emerges as

$$\mathbf{e}^{(1/2)}(\mathbf{k}) \cdot \overline{\left[ \hat{\mathbf{k}} \times \mathbf{e}^{(1/2)}(\mathbf{k}) \right]} = (-1)^{(2/1)} i \sin(2\theta) \sin(\phi).$$

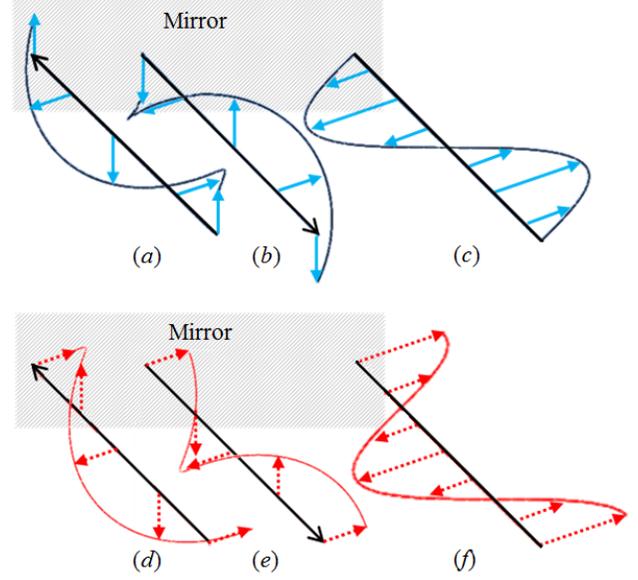

**FIGURE 1.** The electric (top) and magnetic (bottom) field vectors of circularly polarized light under reflection. The interference (c,f) of the input (a,d) and reflected (b,e) beams results in superposition states whose minima and maxima are in different locations for the electric and magnetic fields.

In the same way, for spin-type measures the result will involve either of the following vector cross-products:

$$\mathbf{e}_{(1/2)} \times \overline{\mathbf{e}}_{(1/2)} = \left[ \hat{\mathbf{k}} \times \mathbf{e}^{(1/2)}(\mathbf{k}) \right] \times \overline{\left[ \hat{\mathbf{k}} \times \mathbf{e}^{(1/2)}(\mathbf{k}) \right]}$$
$$= (-1)^{(1/2)} i \sin(2\theta) \sin(\phi) \hat{\mathbf{k}}.$$

Therefore, all helicity-type and spin-type operators obtained via repeated curls of the fields and potentials – as appear in Eqs (6), (7) – emerge with the characteristic dependence on the difference between number operators for optical modes of opposing helicity.

Turning attention to the orbital angular momentum operator, Eq. (2), we use the quantum expansions for the vector potential of a Laguerre-Gaussian (LG) mode as a test case for beams bearing OAM [18]. In the paraxial approximation, the magnetic and transverse electric field vectors can again be determined from the vector potential, the latter now given by

$$\mathbf{A} = \sum_{\substack{\mathbf{k},\eta \\ l,p}} \left( \frac{\hbar}{2\varepsilon_0 c k V} \right)^{\frac{1}{2}} \left\{ \mathbf{e}_{l,p}^{(\eta)}(\mathbf{k}) a^{(\eta)}(\mathbf{k}) f_{l,p}(r) e^{ikz - il\varphi} + h.c. \right\}$$

where $f_{l,p}(r)$ represents the radial distribution of the LG mode with radial number *p* and azimuthal index *l*. In the case of plane waves, the analysis of orbital angular momentum using Eq. (2) gives a vanishing result. When applied to a field of mode $(\mathbf{k},\lambda,l,p)$ the operator becomes

$$L = \hbar \sum_{\mathbf{k},l,p} l\hat{\mathbf{k}} \left[ \hat{N}_{lp}^{(1)}(\mathbf{k}) + \hat{N}_{lp}^{(2)}(\mathbf{k}) \right].$$

Hence, it can be asserted that in the paraxial approximation the separation into spin and orbital angular momentum is equivalent to a division of the optical angular momentum into parts that respectively have a dependence on the difference, and the sum, of number operators for modes of opposite polarization helicity.

To achieve a complete picture of results [13, 19] that have contentiously been described as involving 'superchiral light', it is instructive to consider related pedagogical issues that arise in connection with the reflection of circularly polarized light. At an idealized mirror, the incidence of a photon with wave-vector **k** leads to the emergence of a photon with wave-vector -**k**. The interaction involves no change in photon spin, and therefore there is a reversal of the helicity, signifying the projection of the spin angular momentum onto the direction of propagation [15]. The electric field vector changes sign but, to preserve the right handed triad (**E**, **B**, **k** for the incident field; corresponding quantities for the emergent) it is apparent that the magnetic field does not. Thus, the superposition state of the photon and its reflection near the surface of the mirror is one with a small electric field and a large magnetic field (Fig. 1). Notably, the former never entirely vanishes, due to the exhibition of quantum uncertainty features [12]. With this in mind, consider the potentially circular differential response of a chiral molecule at this location, as might be exhibited in an electronic transition that is both *E1* and *M1* (electric dipole and magnetic dipole) allowed. Engaging the molecule with such an optical state will succeed in suppressing the $E1^2$ rate contribution – one that cannot display enantioselectivity – while permitting potentially larger rate contributions from the *E1-M1* interference, the leading order chiral correction to the absorption rate.

To summarize: In a precise quantum optical formalism, it has been shown that, for both paraxial and non-paraxial light, known measures of electromagnetic phenomena share a commonality: their terms contain an equal number of annihilation and creation operators – a property emerging from their bilinearity in the electric and magnetic fields. Considering the photonic nature of light and a Poincaré sphere representation of the polarization vectors [16], the infinite hierarchy of helicity-type and spin-type measures found by Cameron, Barnett and Yao [15] acquire a physically meaningful form: they all depend on the difference between number operators for optical modes of opposing polarization helicity. Therefore, these measures can never exhibit a value greater than for a purely circularly polarized beam. The reflection of such a beam by a mirror has been discussed and regions of prominent chiral interaction in the resulting interference have been identified.

The authors thank EPSRC for funding this research.